\documentclass[10pt,twocolumn]{article}

\usepackage[T1]{fontenc}
\usepackage[utf8]{inputenc}
\usepackage{mathptmx}            
\usepackage{amsmath,amssymb}
\usepackage{microtype}
\usepackage[letterpaper,top=0.75in,bottom=1in,left=0.625in,right=0.625in]{geometry}
\usepackage{graphicx}
\usepackage{booktabs}
\usepackage{multirow}
\usepackage{makecell}
\usepackage{caption}
\usepackage{subcaption}
\usepackage{titlesec}
\usepackage{cite}
\usepackage{balance}
\usepackage[hidelinks]{hyperref}

\renewcommand{\thesection}{\Roman{section}}

\titleformat{\section}{\centering\normalsize\scshape}{\thesection.}{0.5em}{}
\titlespacing{\section}{0pt}{12pt plus 2pt minus 2pt}{6pt}
\titleformat{\subsection}{\normalsize\itshape}{\Alph{subsection}.}{0.5em}{}
\titlespacing{\subsection}{0pt}{10pt plus 2pt minus 2pt}{4pt}
\titleformat{\subsubsection}[runin]{\normalsize\itshape}{\arabic{subsubsection})}{0.5em}{}

\newcommand{\eg}{e.g.,\ }

\begin{document}

\twocolumn[{%
\begin{center}
{\LARGE Data-driven Video Codec with Implicit Neural\\[2pt] Representations\par}
\vspace{1.4em}
{\large Nishan Khanal\textsuperscript{*}, Saugat Neupane\textsuperscript{*}, Abhinav Chalise\textsuperscript{*}, Nimesh Gopal Pradhan\textsuperscript{*}, Dinesh Baniya Kshatri\textsuperscript{*}\par}
\vspace{0.4em}
{\itshape Department of Electronics and Computer Engineering, Thapathali Campus\par}
{\itshape Institute of Engineering, Tribhuvan University, Kathmandu, Nepal\par}
\vspace{0.5em}
{\small khanal.nishan28@gmail.com, saugatn3@gmail.com, chalisezabhinav@gmail.com,\par}
{\small nimeshgpradhan@gmail.com, dinesh@ioe.edu.np\par}
\vspace{1.8em}
\end{center}
}]

\noindent\textbf{\small Abstract---A conventional codec stores a video as compressed pixel data. We instead store the video, together with its audio track, as the weights of a single sinusoidal representation network (SIREN) that maps space--time coordinates to RGB values and audio amplitudes. The network uses separate audio and video initialization layers, a stack of shared fully connected hidden layers, and three output branches: one for video and two Siamese audio branches whose disagreement is used to estimate and subtract residual noise. The overfitted teacher network is then compressed by response-based knowledge distillation into a smaller student, followed by 16-bit symmetric weight quantization and lossless LZMA2 (xz) encoding. On a 6.08~MiB test video, the quantized student reaches a video PSNR of 28.72~dB with SSIM of 0.75, and an audio PSNR of 24.18~dB with a log spectral distance of 10.69~dB, while the pipeline shrinks the representation from 9.05~MiB to 2.33~MiB, an overall compression ratio of 2.61. A bit-width sweep from 1-bit to 32-bit quantization shows that reconstruction quality saturates at 16 bits. We compare against H.264, HEVC, and MP3, report where the approach falls short of them, and describe a browser-based prototype that trains, transfers, and decodes these models over WebRTC.}

\vspace{0.6em}
\vspace{0.8em}

\section{Introduction}

Video accounts for most of the data stored and moved on the modern internet, and the pressure on codecs keeps growing as 4K and 8K formats become ordinary. The standards that carry this load, H.264/AVC \cite{kalva2006h264} and H.265/HEVC \cite{pourazad2012hevc}, compress pixel data directly: they remove spatial and temporal redundancy with transforms, motion prediction, and entropy coding. They are fast and mature, but they operate on a fixed representation of the signal --- a grid of pixels --- and their efficiency gains between generations have come at the cost of considerable encoder complexity.

Implicit neural representations (INRs) take a different route. Instead of storing samples of a signal, an INR trains a neural network to \emph{be} the signal: the network maps a coordinate (\eg pixel position and frame time) to the signal value at that coordinate, and the trained weights become the stored representation. Sitzmann et al.\ showed that multilayer perceptrons with sine activations (SIRENs) fit images, audio, and video far better than ReLU networks of the same size \cite{sitzmann2020siren}, and NeRV demonstrated that video compression can be recast as \emph{model} compression once the video lives inside a network \cite{chen2021nerv}. Since then, INR-based codecs have improved steadily \cite{gomes2023entropy,zhang2024boosting}, and Siamese SIREN extended the idea to audio, using two lightly different network heads to estimate and remove reconstruction noise \cite{lanzendorfer2023siamese}.

Almost all of this work treats video and audio separately. A video INR stores the frames; the soundtrack is left to a conventional audio codec or to a second network. NeRVA \cite{choudhury2024nerva} is a notable exception, but it operates frame-wise and mixes convolutional blocks with MLPs. To our knowledge, no prior work represents both modalities \emph{sample-wise} in one plain MLP, where a single set of shared hidden layers serves both the pixel function and the amplitude function.

This paper asks whether that unified representation is practical, and what it costs. We make four contributions:

\begin{itemize}
\item \textbf{A unified audio--video SIREN.} One MLP takes normalized space--time coordinates $[(x,y,t),T]$ and returns RGB values and audio amplitudes. Audio and video enter through separate initialization layers (they need very different first-layer weight ranges), share five hidden layers, and exit through one video branch and two Siamese audio branches used for noise estimation.
\item \textbf{A compression pipeline for the representation.} Response-based knowledge distillation shrinks the 2.37M-parameter teacher to a 1.30M-parameter student; 16-bit symmetric quantization halves the stored size again; LZMA2 encoding removes what redundancy remains. The full pipeline takes the representation of one test video from 9.05~MiB to 2.33~MiB.
\item \textbf{A quantization study.} We quantize the student from 1 to 32 bits and measure PSNR, SSIM, and signal-to-quantization-noise ratio. Quality collapses below 10 bits and saturates at 16, which fixes our operating point.
\item \textbf{An end-to-end prototype.} A FastAPI backend trains and decodes models; browsers exchange the compressed weights peer-to-peer over WebRTC.
\end{itemize}

We also report the negative results plainly. The teacher network has a fixed size of about 9~MiB regardless of input, so short clips are \emph{expanded} rather than compressed, and on most test content the codec trails H.264 and HEVC at comparable quality. The interesting property is that the stored size is constant in the video length and resolution --- the codec gets relatively better as the input gets larger --- and that a single cheap MLP carries both modalities at all.

\section{Related Work}

\textbf{Conventional codecs.} MPEG-1 Layer~3 established perceptual audio coding, using psychoacoustic models and Huffman coding to discard inaudible detail \cite{shlien1994mpeg}. H.264/AVC brought intra-picture prediction, variable block sizes, multiple reference frames, and in-loop deblocking to video \cite{kalva2006h264}; HEVC roughly doubled its compression efficiency through more flexible block partitioning and better prediction \cite{pourazad2012hevc}. These standards are the baselines any learned codec must face.

\textbf{Learned video compression.} Li et al.\ replaced residue coding with conditional coding, letting the encoder and decoder condition on learned feature-domain context, and reported 26\% bitrate savings over x265 \cite{li2021dcvc}; a later version with richer temporal and spatial contexts surpassed the ECM reference software \cite{li2023diverse}. Liu et al.\ compressed video in the latent space of a GAN-trained autoencoder with a ConvLSTM predictor \cite{liu2021latent}. Panneerselvam et al.\ combined CNN-based duplicate-frame removal, GAN-detected frame differences, and SVD to halve file sizes with about 10\% quality loss \cite{panneerselvam2022effective}.

\textbf{Implicit neural representations.} SIREN showed that sine activations, with a principled initialization, let coordinate MLPs represent signals and their derivatives accurately \cite{sitzmann2020siren}. NeRV turned video coding into model fitting: a network maps a frame index to an RGB frame, and standard model-compression tools then act as the video codec \cite{chen2021nerv}. Gomes et al.\ added entropy-constrained training so that rate and distortion are optimized jointly, removing the need for post-hoc quantization \cite{gomes2023entropy}, and Zhang et al.\ boosted NeRV-like representations with conditional decoders and consistent entropy minimization \cite{zhang2024boosting}. On the audio side, Siamese SIREN found that two lightly perturbed SIREN heads make reconstruction noise estimable: the difference of the two outputs approximates the noise, which a spectral-gating step then removes \cite{lanzendorfer2023siamese}. NeRVA represents video and audio jointly, but frame-wise and with convolutional blocks \cite{choudhury2024nerva}. KD-INR applied knowledge distillation to INRs for time-varying volumetric data, aggregating per-timestep models into one compact network \cite{han2024kdinr}. Quantization tooling for all of these is well developed in mainstream frameworks \cite{pytorch_quant}.

\textbf{Gap.} What is missing is a sample-wise, MLP-only representation in which one network jointly stores pixels and audio amplitudes and is then compressed as a unit. That is the configuration this paper builds and measures.

\section{Method}

\subsection{SIREN Preliminaries}

A SIREN layer applies a sine to an affine map,
\begin{equation}
x_{i+1} = \sin(W_i x_i + b_i),
\label{eq:siren}
\end{equation}
where $W_i$ and $b_i$ are the weights and bias of layer $i$. Because the sine and all of its derivatives are smooth and periodic, the network can fit high-frequency content that ReLU MLPs smooth away, and training remains stable at depth \cite{sitzmann2020siren}. Weights are initialized uniformly in $[-\sqrt{c/n},\sqrt{c/n}]$ for fan-in $n$, which keeps activations arcsine-distributed from layer to layer, and the first layer is scaled by a frequency factor $\omega_0$ so that $\sin(\omega_0 W x + b)$ spans several periods over the normalized input range. We use the standard $\omega_0 = 30$.

\subsection{Unified Audio--Video Architecture}

\begin{figure*}[!t]
\centering
\includegraphics[width=0.92\textwidth]{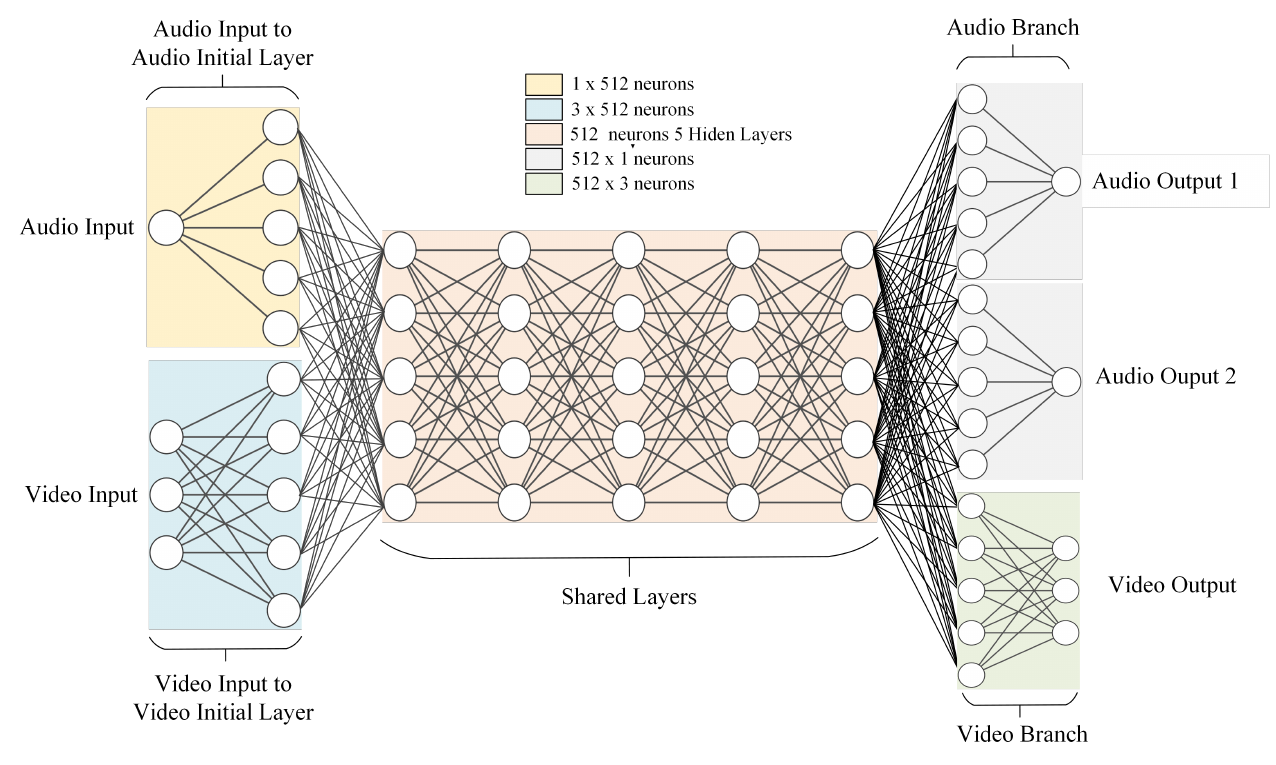}
\caption{The unified SIREN. Audio timestep $T$ and space--time coordinates $(x,y,t)$ enter through separate initialization layers, pass through shared sine-activated hidden layers, and exit through one video branch and two Siamese audio branches.}
\label{fig:arch}
\end{figure*}

Figure~\ref{fig:arch} shows the network. The input is a normalized coordinate tuple $[(x,y,t),T]$: pixel position $(x,y)$, frame index $t$, and audio timestep $T$, each scaled to $[-1,1]$ or $[0,1]$ as appropriate. The output is the RGB value at $(x,y,t)$ and two audio amplitudes at $T$.

Two design points matter here. First, audio and video need \emph{separate initialization layers}. Audio is far richer in high frequencies than video, and we found no single first-layer weight range that serves both: the audio branch needs weights initialized in $[-25,25]$ to capture its spectrum, while the video branch needs the much narrower $[-2/3,\,2/3]$ --- wider ranges produced grainy frames and wrong colors, narrower ones muffled the audio. So each modality gets its own first layer, and their outputs are concatenated before the shared stack.

\begin{figure}[!t]
\centering
\includegraphics[width=\columnwidth]{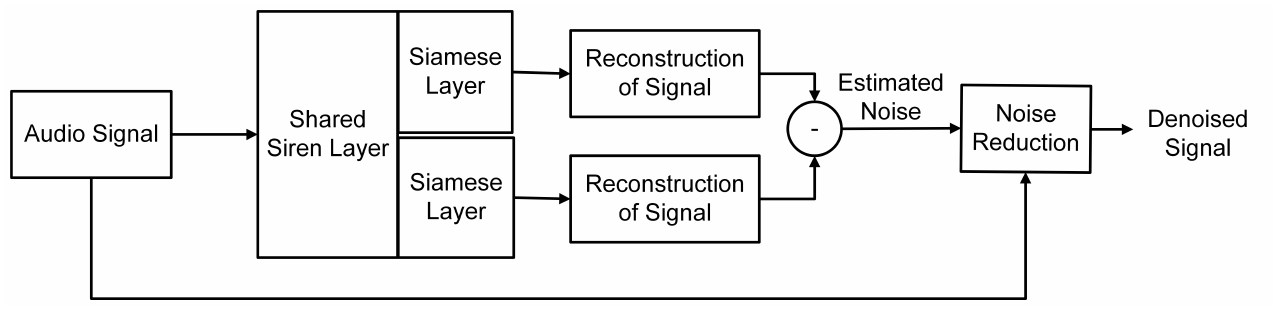}
\caption{Siamese SIREN principle: two branches infer the same audio; their difference estimates the noise.}
\label{fig:siamese}
\end{figure}

Second, the audio output is duplicated. Following Siamese SIREN \cite{lanzendorfer2023siamese}, two identical linear audio branches with different weight draws produce two estimates $f_1$ and $f_2$ of the same amplitude. Their difference,
\begin{equation}
\widehat{\text{noise}} = f_1 - f_2,
\label{eq:noise}
\end{equation}
estimates the reconstruction noise and drives the spectral denoising step of Section~\ref{sec:postproc}. Figure~\ref{fig:siamese} sketches the idea.

The teacher uses five shared hidden layers of 512 neurons; the student, three. In the student, all non-hidden layers are reduced to 300 neurons while hidden layers stay at 512. All layers use sine activations except the three output branches, which are linear. Table~\ref{tab:params} gives the parameter breakdown: 2{,}369{,}029 trainable parameters for the teacher and 1{,}298{,}029 for the student.

\begin{table}[!t]
\caption{Trainable Parameters in the SIREN Models}
\label{tab:params}
\centering
\scriptsize
\setlength{\tabcolsep}{3pt}
\begin{tabular}{@{}lccrr@{}}
\toprule
& \multicolumn{2}{c}{Dimensions} & \multicolumn{2}{c}{Total} \\
\cmidrule(lr){2-3}\cmidrule(l){4-5}
Layer & Teacher & Student & Teacher & Student \\
\midrule
Audio initial   & $1\times512$ & $1\times300$ & 1{,}024 & 600 \\
Video initial   & $3\times512$ & $3\times300$ & 2{,}048 & 1{,}200 \\
Shared (first)  & $1024\times512$ & $600\times512$ & 524{,}800 & 307{,}712 \\
Shared (rest)   & $4\times(512\times512)$ & $2\times(512\times512)$ & 1{,}050{,}624 & 525{,}312 \\
Video branch    & $512\times512,\,512\times3$ & $512\times300,\,300\times3$ & 264{,}195 & 154{,}803 \\
Audio branch 1  & $512\times512,\,512\times1$ & $512\times300,\,300\times1$ & 263{,}169 & 154{,}201 \\
Audio branch 2  & $512\times512,\,512\times1$ & $512\times300,\,300\times1$ & 263{,}169 & 154{,}201 \\
\midrule
Total & & & 2{,}369{,}029 & 1{,}298{,}029 \\
\bottomrule
\end{tabular}
\end{table}

\subsection{Codec Pipeline}

\begin{figure}[!t]
\centering
\includegraphics[width=0.98\columnwidth]{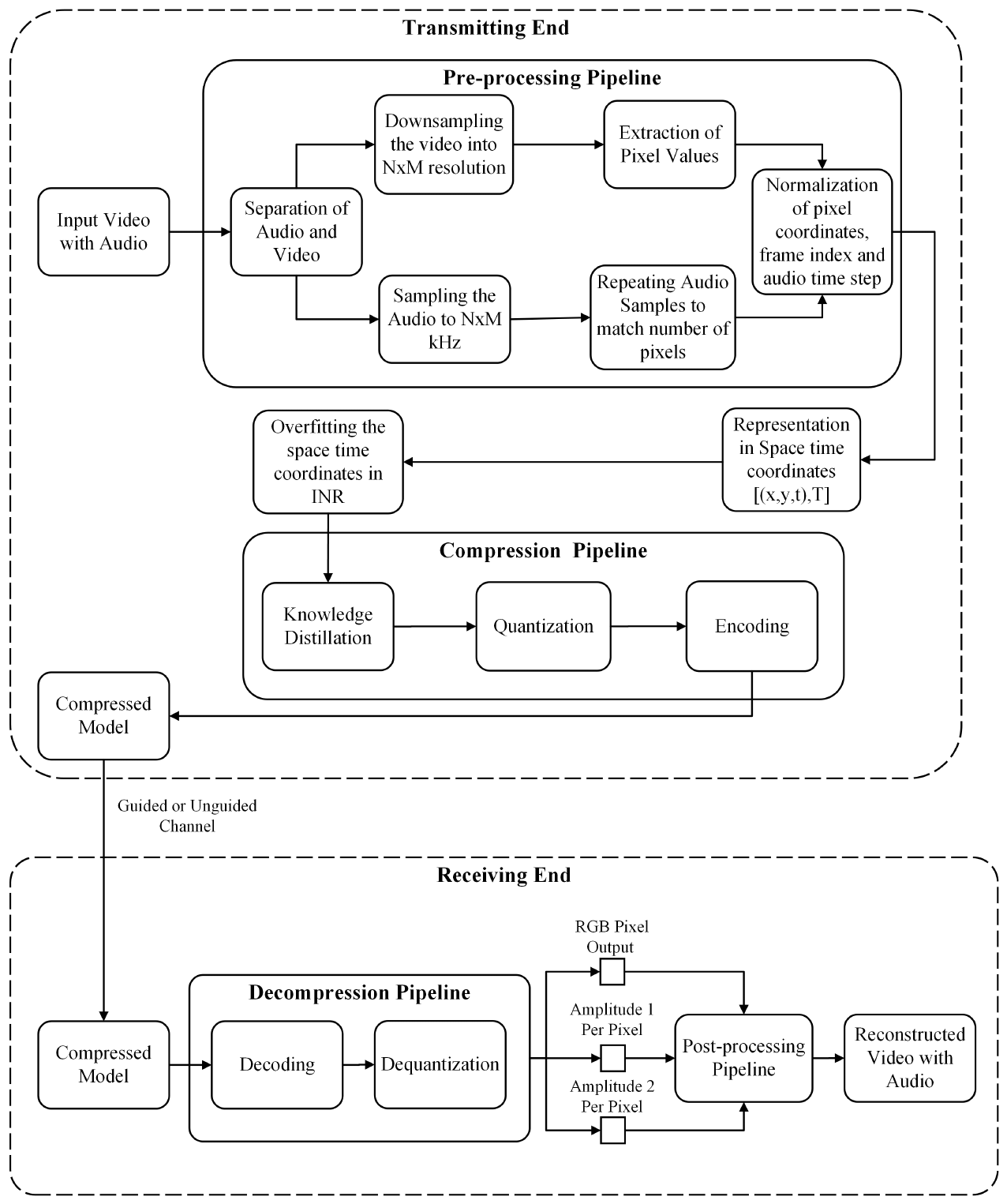}
\caption{System block diagram. The transmitting end overfits the INR and compresses it by distillation, quantization, and encoding; the receiving end decodes, dequantizes, and runs inference to reconstruct the video with audio.}
\label{fig:system}
\end{figure}

Figure~\ref{fig:system} shows the codec end to end. Encoding a video means training a network on it; decoding means running inference over the coordinate grid. The stages are as follows.

\subsubsection{Pre-processing}
\begin{figure}[!t]
\centering
\includegraphics[width=0.9\columnwidth]{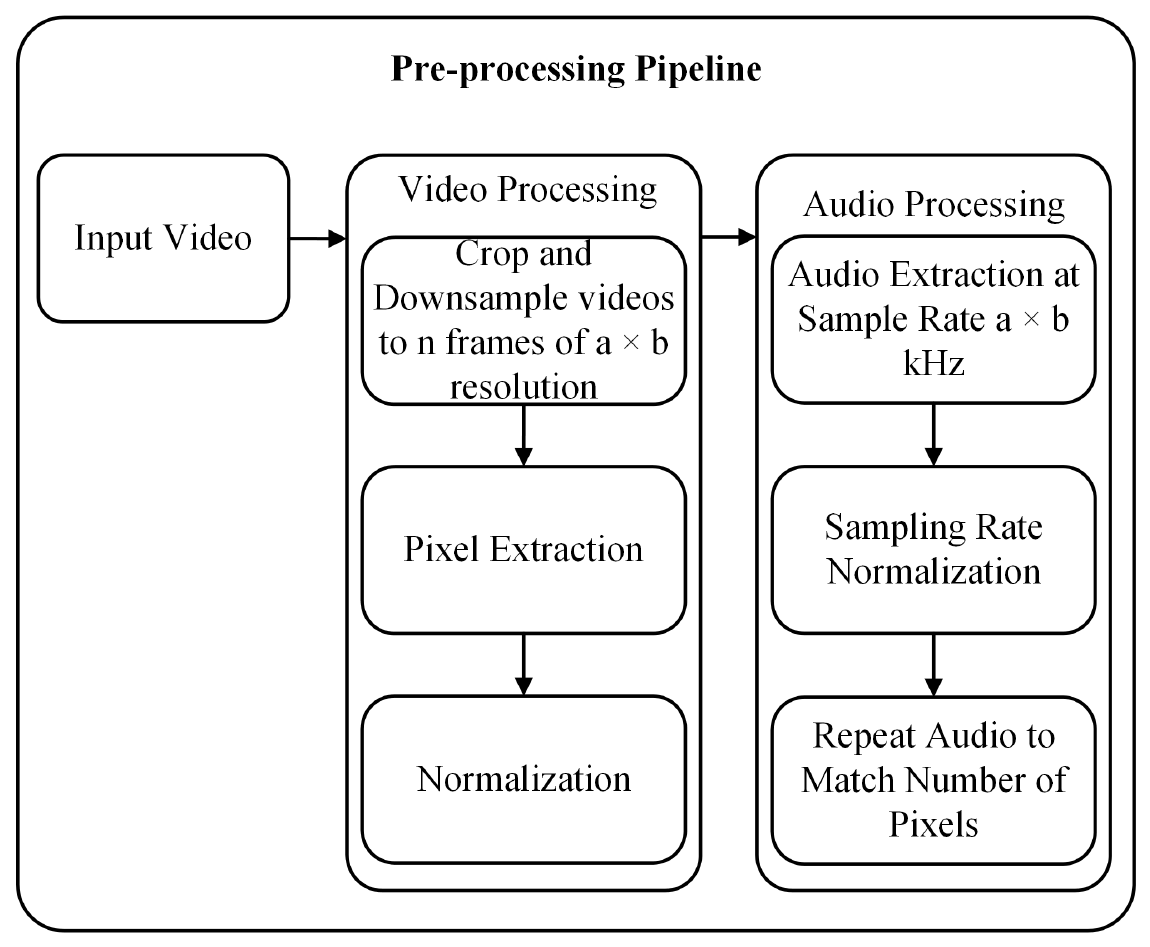}
\caption{Pre-processing pipeline.}
\label{fig:preproc}
\end{figure}

The input container is split into video and audio streams (Fig.~\ref{fig:preproc}). Frames are cropped and downsampled to a fixed $a\times b$ resolution and their pixels normalized to $[0,1]$; audio is resampled to a rate matched to the pixel count and normalized by its peak amplitude. Because one frame contains far more pixels than there are audio samples in its display interval, audio samples are repeated until the two streams have equal length; the repetition is removed after decoding. Pixel coordinates, frame indices, and audio timesteps are then normalized, producing training pairs of coordinates and target values.

\subsubsection{Overfitting and compression}
The teacher is trained to overfit the coordinate-to-value mapping (Section~\ref{sec:training}). The compression pipeline then applies, in order: knowledge distillation into the student, 16-bit symmetric quantization of weights and biases, and lossless LZMA2 encoding with \texttt{xz}. The result is the transmitted bitstream.

\subsubsection{Decoding and post-processing}
\label{sec:postproc}
\begin{figure}[!t]
\centering
\includegraphics[width=0.98\columnwidth]{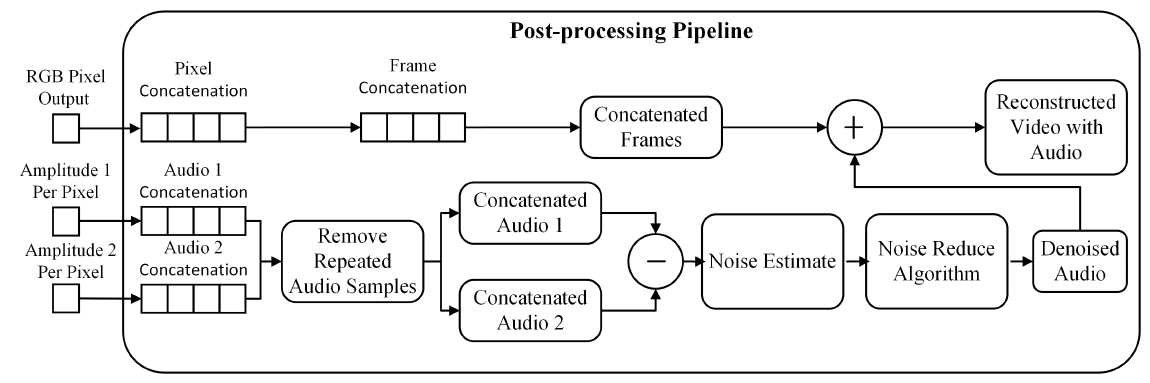}
\caption{Post-processing pipeline.}
\label{fig:postproc}
\end{figure}

The receiver decodes the LZMA2 stream, dequantizes weights using the stored per-layer scale factors, and evaluates the network over the full coordinate grid. RGB outputs are reassembled into frames (Fig.~\ref{fig:postproc}). The two audio outputs are concatenated, their sample repetition removed, and their difference used as the noise estimate of \eqref{eq:noise}. The denoiser computes the spectrogram statistics of the noise estimate (per-band mean and standard deviation), forms a threshold $\mu + k\sigma$ per frequency band, masks the audio spectrogram below the threshold, smooths the mask with a Gaussian filter, and inverts the masked STFT to obtain the final audio.

\subsection{Training Objective}
\label{sec:training}

The teacher minimizes a weighted sum of per-modality reconstruction losses,
\begin{equation}
\mathcal{L}_{\text{combined}} = \lambda_{\text{audio}}\,\mathcal{L}_{\text{audio}} + \lambda_{\text{video}}\,\mathcal{L}_{\text{video}},
\end{equation}
with
\begin{align}
\mathcal{L}_{\text{audio}} &= \frac{1}{2}\sum_{k=1}^{2}\int_{\Omega_a}\!\bigl\|\Phi_{a,k}(T) - f_a(T)\bigr\|\,dT, \\
\mathcal{L}_{\text{video}} &= \int_{\Omega_v}\bigl\|\Phi_v(x,t) - f_v(x,t)\bigr\|\,dx\,dt,
\end{align}
where $\Phi_{a,1},\Phi_{a,2}$ are the two audio branch outputs, $\Phi_v$ the video output, and $f_a$, $f_v$ the ground-truth amplitude and RGB functions. In practice the integrals are MSE sums over the sampled coordinates.

\begin{figure}[!t]
\centering
\includegraphics[width=0.95\columnwidth]{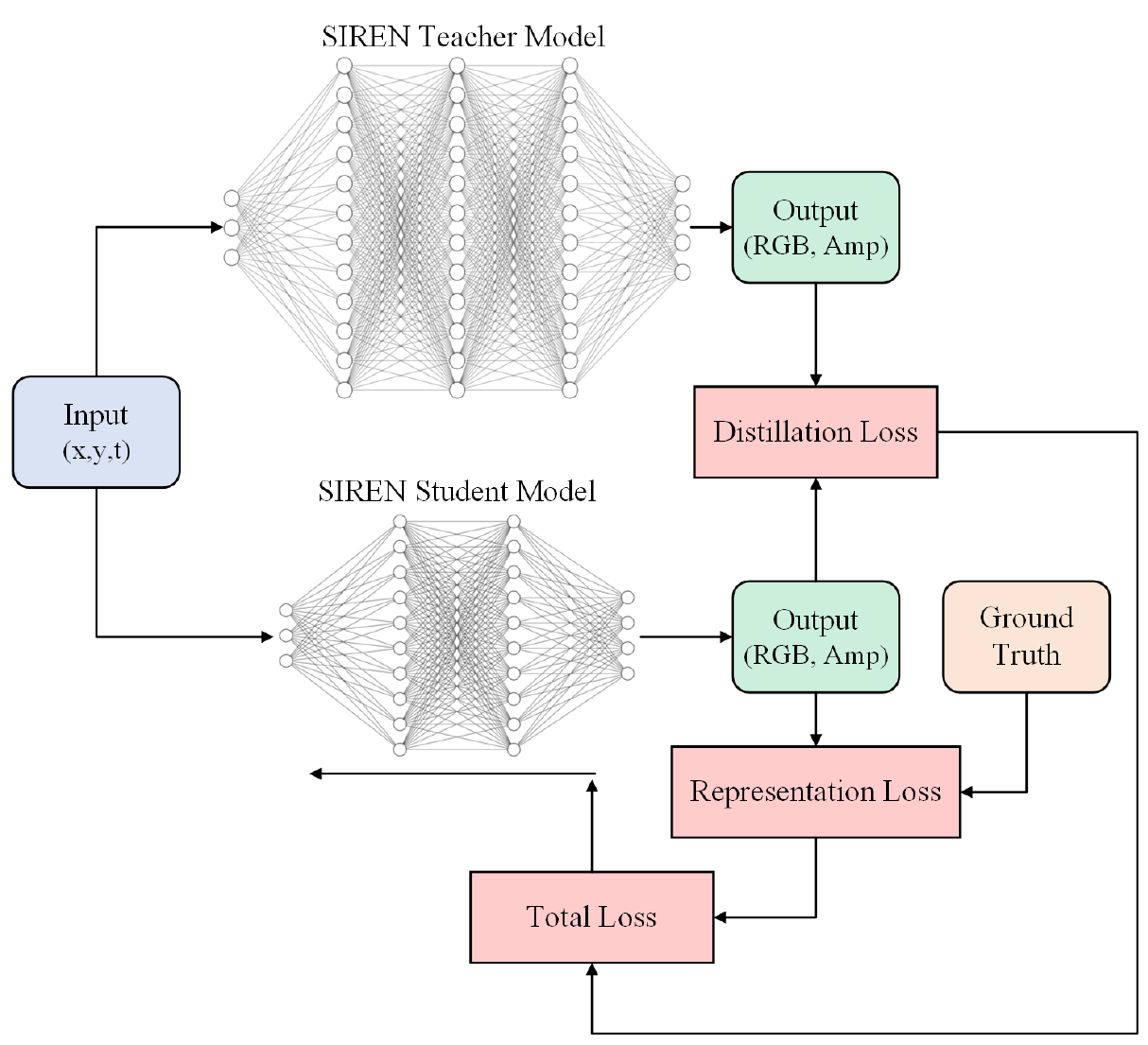}
\caption{Response-based knowledge distillation. The student is supervised by the ground truth (hard loss) and by the teacher's outputs (soft loss).}
\label{fig:kd}
\end{figure}

The student is trained by response-based knowledge distillation (Fig.~\ref{fig:kd}): it sees both the ground truth and the frozen teacher's outputs,
\begin{equation}
\mathcal{L}_{\text{KD}} = \alpha\,\mathcal{L}_{\text{hard}} + (1-\alpha)\,\mathcal{L}_{\text{soft}},
\end{equation}
where the hard loss compares student outputs with ground truth and the soft loss compares them with teacher outputs. Both are split evenly between video and audio, and the audio half is split again between the two Siamese branches:
\begin{equation}
\mathcal{L}_{\bullet} = 0.5\,\mathcal{L}_{\text{video},\bullet}
+ 0.5\bigl(0.5\,\mathcal{L}_{\text{audio},\bullet} + 0.5\,\mathcal{L}_{\text{audio(siam)},\bullet}\bigr).
\end{equation}

\subsection{Quantization and Encoding}

Weights are quantized symmetrically to 16-bit integers,
\begin{equation}
w' = \operatorname{round}\!\left(\frac{w}{s}\right) + z,
\label{eq:quant}
\end{equation}
with zero-point $z=0$ and a per-layer scale $s$ chosen from the layer's weight range; scale factors are stored with the model for dequantization. Section~\ref{sec:quantstudy} justifies the choice of 16 bits empirically. The quantized checkpoint is finally compressed with LZMA2 via \texttt{xz}, which is lossless, so it changes no quality metric --- only the file size.

\section{Experimental Setup}

\subsection{Dataset}

We evaluate on five short videos with audio, stored uncompressed in AVI/WAV so that no prior codec biases the ground truth. Table~\ref{tab:dataset} lists their characteristics; all are $448\times256$. Videos~1 and~2 have temporally uncorrelated frames, chosen to stress the model when consecutive frames change completely; their audio is a violin melody (1.6~Hz--4.4~kHz) and a Nepali male speaker (0.3~Hz--5.7~kHz), respectively. Videos~3--5 have correlated frames: a man dancing to music (0.1~Hz--15.5~kHz), people dancing to funk (0.4~Hz--10.3~kHz), and an English speech (1.2~Hz--10.5~kHz). Audio sample rates are set high so that fewer repetitions are needed to match the pixel count during pre-processing.

\begin{table}[!t]
\caption{Video File Characteristics}
\label{tab:dataset}
\centering
\footnotesize
\begin{tabular}{@{}clcccc@{}}
\toprule
SN & Video & FPS & Sample rate (Hz) & Size (MiB) & Duration (s) \\
\midrule
1 & Video 1 & 1  & 114{,}688 & 1.92 & 5 \\
2 & Video 2 & 1  & 96{,}000  & 5.33 & 10 \\
3 & Video 3 & 10 & 114{,}688 & 6.08 & 3 \\
4 & Video 4 & 2  & 114{,}688 & 13.6 & 25 \\
5 & Video 5 & 24 & 114{,}688 & 12.4 & 3 \\
\bottomrule
\end{tabular}
\end{table}

\subsection{Training Details}

Everything is implemented in PyTorch and trained on a laptop with an Intel Core i9-13980HX, 24~GiB of RAM, and an NVIDIA RTX~4090 laptop GPU with 16~GiB of VRAM. Table~\ref{tab:hyper} lists the hyperparameters. The teacher trains for 7{,}000 epochs at a learning rate of $10^{-5}$; the student for 6{,}000 epochs, starting at $10^{-4}$ for the first 1{,}000 epochs and dropping to $10^{-5}$ for the rest. Both use Adam and $\omega_0=30$, and the checkpoint with the lowest loss is kept. Training continues well past the point where the loss flattens because the perceptual metrics (PSNR, SSIM, LPIPS, LSD) keep improving after the raw MSE has nearly converged.

\begin{table}[!t]
\caption{Hyperparameters for the Teacher and Student Models}
\label{tab:hyper}
\centering
\footnotesize
\begin{tabular}{@{}lcc@{}}
\toprule
Hyperparameter & Teacher & Student \\
\midrule
Epochs & 7{,}000 & 6{,}000 \\
$\omega_0$ & \multicolumn{2}{c}{30} \\
First-layer weights, audio ($\beta_a$) & \multicolumn{2}{c}{$(-25,\,25)$} \\
First-layer weights, video ($\beta_v$) & \multicolumn{2}{c}{$(-2/3,\,2/3)$} \\
Learning rate ($\eta$) & $1\times10^{-5}$ & $10^{-4}$ (first 1{,}000 ep.), \\
 & & then $10^{-5}$ \\
Optimizer & \multicolumn{2}{c}{Adam} \\
\bottomrule
\end{tabular}
\end{table}

\subsection{Metrics}

Video quality is measured with PSNR, SSIM, and LPIPS against the source frames, and with SQNR when comparing a quantized model with its unquantized counterpart. Audio quality is measured with PSNR, log spectral distance (LSD), and ViSQOL, an objective estimate of perceived listening quality on a 1--5 scale. File sizes and compression ratios (source size divided by stored model size) complete the picture.

\section{Results and Analysis}

\subsection{Training Behavior}

\begin{figure}[!t]
\centering
\begin{subfigure}[b]{\columnwidth}
\includegraphics[width=\textwidth]{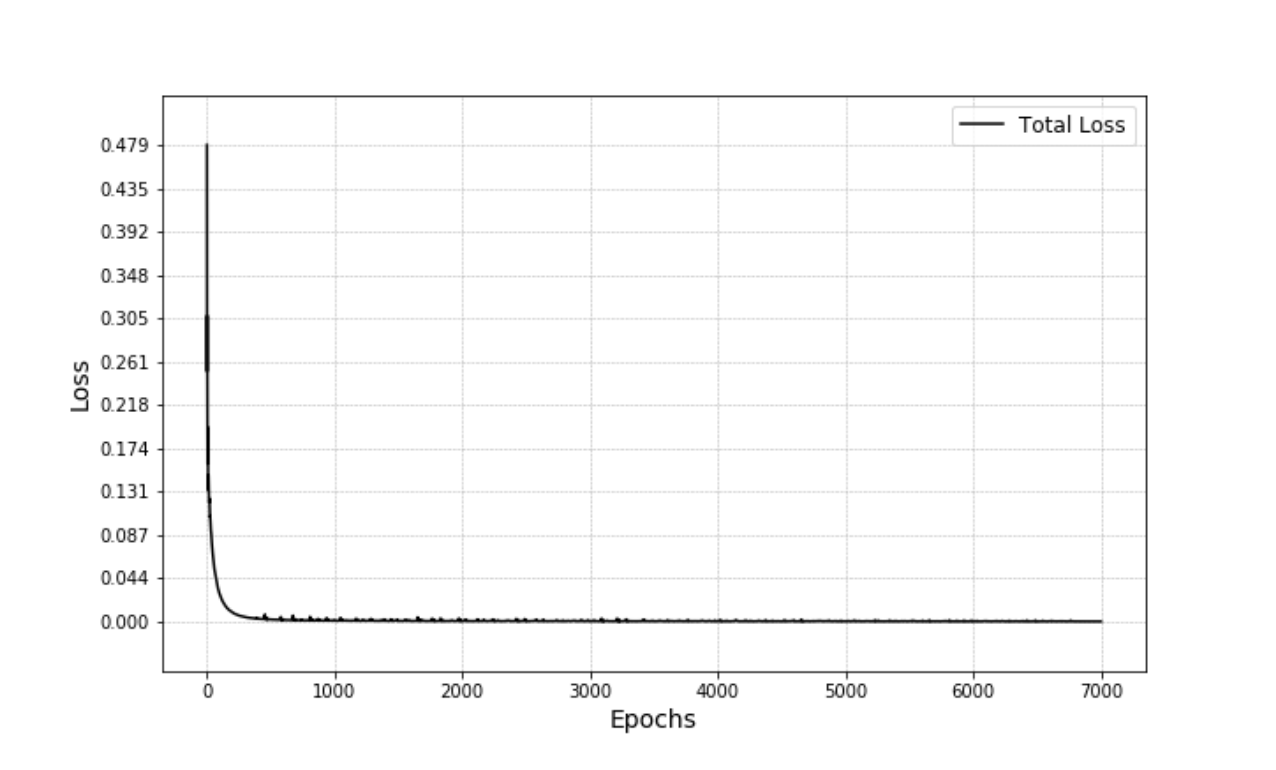}
\caption{Teacher, Video 1.}
\end{subfigure}\\[2pt]
\begin{subfigure}[b]{\columnwidth}
\includegraphics[width=\textwidth]{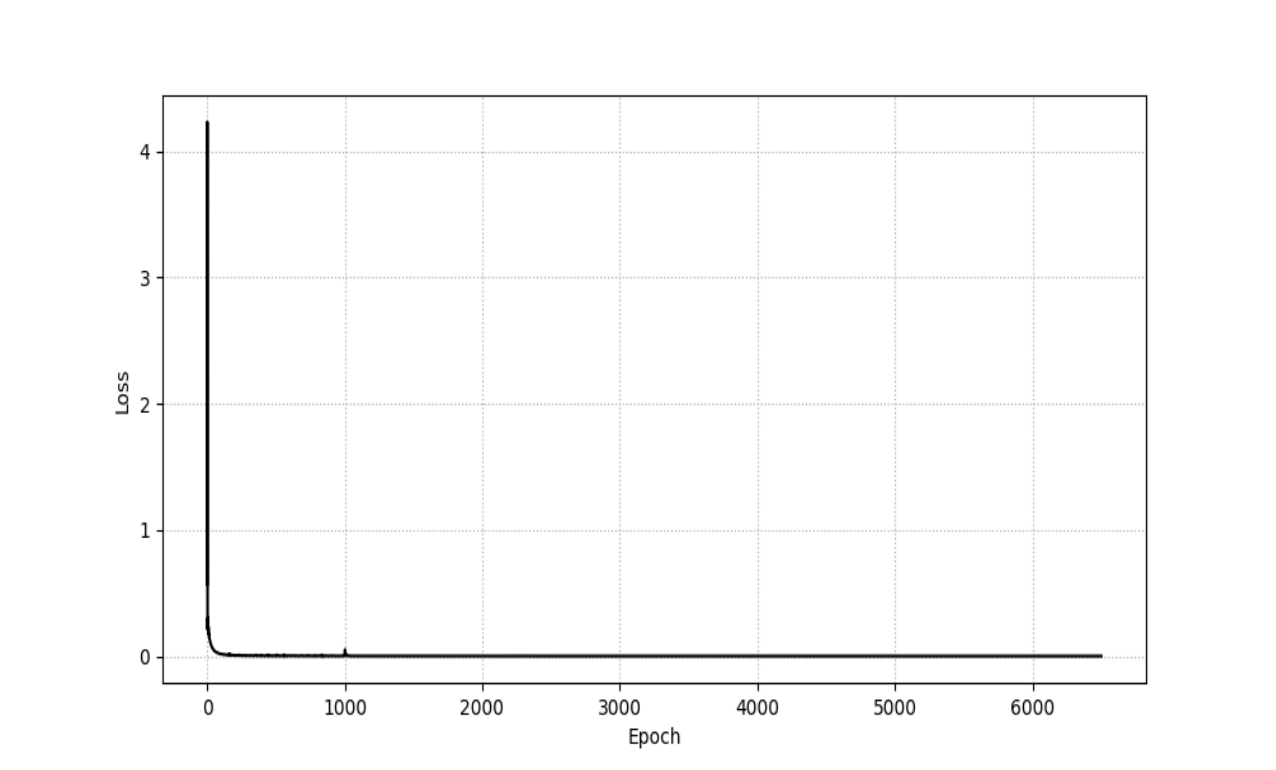}
\caption{Student, Video 1.}
\end{subfigure}
\caption{Training loss curves. Both models drop steeply within the first few hundred epochs and then improve slowly; perceptual metrics keep improving after the loss flattens.}
\label{fig:loss}
\end{figure}

Figure~\ref{fig:loss} shows representative loss curves. The teacher's loss falls steeply and is near zero around epoch 500; the student stabilizes even earlier, near epoch 200. The long tail of training is not wasted: PSNR, SSIM, LPIPS, and LSD continue to improve while the loss changes only in its low digits, which is why we train to 7{,}000 and 6{,}000 epochs rather than stopping at loss convergence.

\subsection{Teacher Model Quality}

\begin{table}[!t]
\caption{Video Metrics of the Teacher Model}
\label{tab:teachervideo}
\centering
\footnotesize
\setlength{\tabcolsep}{3pt}
\begin{tabular}{@{}lcccccc@{}}
\toprule
Video & PSNR (dB) & LPIPS & SSIM & Orig.\ (KiB) & Model (KiB) & Ratio \\
\midrule
Video 1 & 42.69 & 0.05 & 0.98 & 1{,}972  & \multirow{5}{*}{9{,}260} & 0.21 \\
Video 2 & 35.08 & 0.17 & 0.92 & 5{,}462  & & 0.58 \\
Video 3 & 21.88 & 0.39 & 0.67 & 6{,}235  & & 0.67 \\
Video 4 & 26.89 & 0.41 & 0.72 & 14{,}018 & & 1.50 \\
Video 5 & 32.88 & 0.21 & 0.92 & 12{,}781 & & 1.37 \\
\bottomrule
\end{tabular}
\end{table}

\begin{table}[!t]
\caption{Audio Metrics of the Teacher Model}
\label{tab:teacheraudio}
\centering
\footnotesize
\begin{tabular}{@{}lccc@{}}
\toprule
Video & PSNR (dB) & LSD (dB) & ViSQOL \\
\midrule
Video 1 & 57.57 & 4.50 & 3.54 \\
Video 2 & 57.09 & 4.28 & 3.35 \\
Video 3 & 62.50 & 7.60 & 4.51 \\
Video 4 & 64.33 & 8.15 & 3.58 \\
Video 5 & 46.60 & 6.89 & 2.66 \\
\bottomrule
\end{tabular}
\end{table}

Tables~\ref{tab:teachervideo} and~\ref{tab:teacheraudio} report teacher quality. Video PSNR spans 21.88~dB (Video~3, fast dance motion) to 42.69~dB (Video~1, static frames), with SSIM between 0.67 and 0.98. Audio reconstruction is strong across the board --- 46.6 to 64.3~dB PSNR, with ViSQOL up to 4.51 on Video~3.

The model column exposes the codec's defining property: the stored size is 9{,}260~KiB for \emph{every} video, because the network, not the content, sets the size. For Videos~1--3 that means a compression ratio below 1 --- the ``compressed'' file is larger than the source. For the two largest inputs the ratio climbs past 1.3. The size floor is a fixed cost that only pays off once the input is big enough, and everything in Section~\ref{sec:compress} is about lowering that floor.

\subsection{Audio Fidelity}

\begin{figure}[!t]
\centering
\includegraphics[width=0.88\columnwidth]{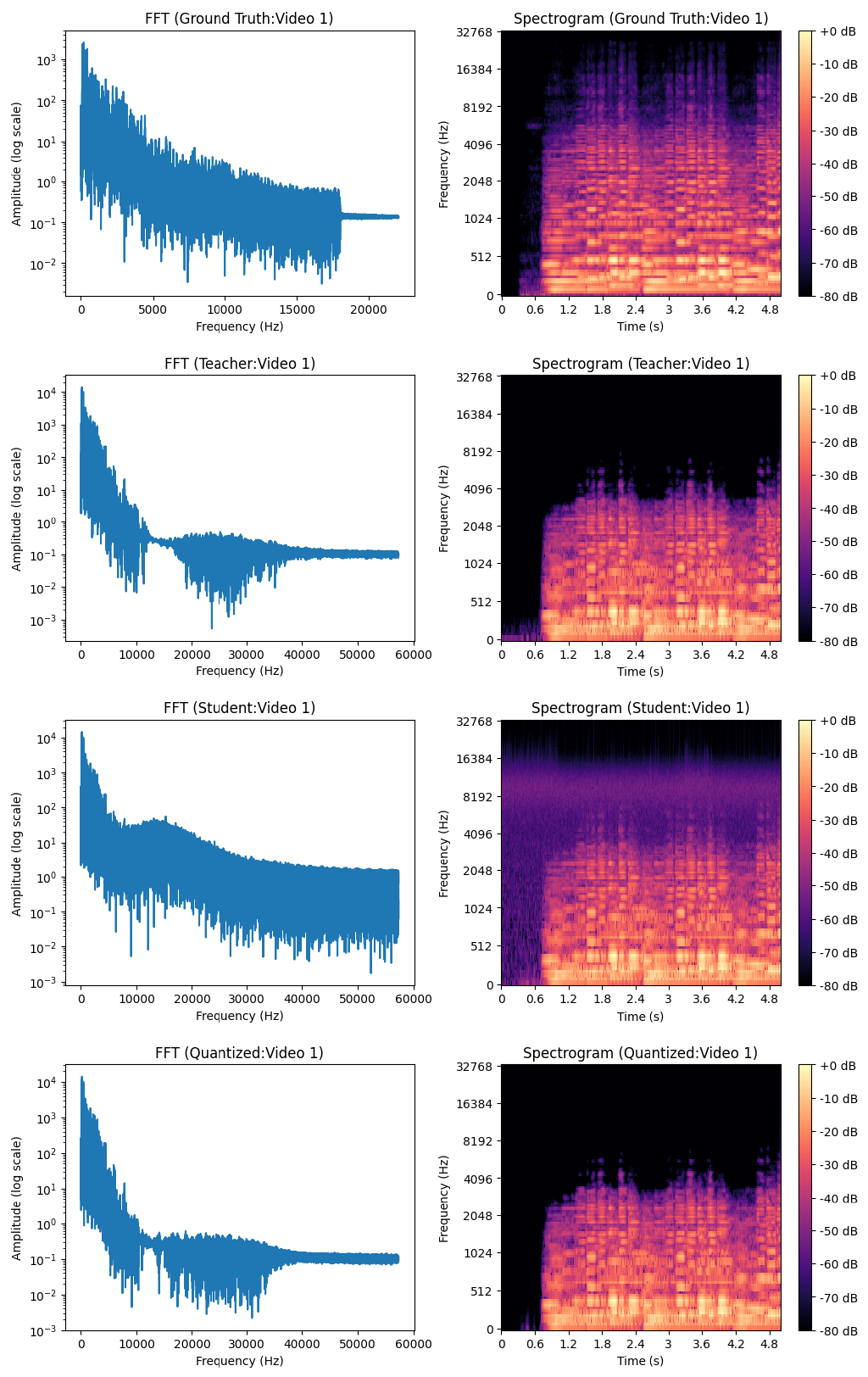}
\caption{FFT and spectrogram comparison of ground-truth and predicted audio for Video 1.}
\label{fig:fft}
\end{figure}

Figure~\ref{fig:fft} compares the ground-truth and reconstructed audio of Video~1 in the frequency and time--frequency domains. The FFT envelopes match closely with small deviations, and the spectrograms are nearly identical; the same holds for the other four videos. The one systematic artifact appears in Videos~2 and~5, where the spectral-gating denoiser misclassifies quiet parts of the signal as noise and removes them --- visible as small gaps in the spectrogram. The noise estimate from the Siamese branches is what makes the denoiser work at all, but its threshold is a blunt instrument on speech pauses.

\subsection{Distillation and 16-bit Quantization}
\label{sec:compress}

\begin{table}[!t]
\caption{Student Model Before and After 16-bit Quantization, Video 1}
\label{tab:quantv1}
\centering
\footnotesize
\begin{tabular}{@{}lcc@{}}
\toprule
Metric & Student & 16-bit quantized \\
\midrule
PSNR, frames (dB) & 35.5525 & 35.3520 \\
SSIM              & 0.9194  & 0.9189 \\
LPIPS             & 0.0794  & 0.0848 \\
PSNR, audio (dB)  & 18.7929 & 18.8487 \\
LSD (dB)          & 4.5930  & 4.8484 \\
SQNR (dB)         & ---     & 39.214 \\
File size (MiB)   & 4.96    & 2.48 \\
Compression ratio & 0.38    & 0.77 \\
\bottomrule
\end{tabular}
\end{table}

\begin{table}[!t]
\caption{Student Model Before and After 16-bit Quantization, Video 3}
\label{tab:quantv3}
\centering
\footnotesize
\begin{tabular}{@{}lcc@{}}
\toprule
Metric & Student & 16-bit quantized \\
\midrule
PSNR, frames (dB) & 28.8588 & 28.7153 \\
SSIM              & 0.7579  & 0.7529 \\
LPIPS             & 0.4039  & 0.4126 \\
PSNR, audio (dB)  & 24.1992 & 24.1786 \\
LSD (dB)          & 10.7038 & 10.6928 \\
File size (MiB)   & 4.96    & 2.48 \\
Compression ratio & 1.22    & 2.45 \\
\bottomrule
\end{tabular}
\end{table}

Tables~\ref{tab:quantv1} and~\ref{tab:quantv3} show the student before and after quantization on Videos~1 and~3. Distillation cuts the checkpoint from 9.05~MiB to 4.96~MiB, and quantization halves it to 2.48~MiB. The quality cost is small: on Video~1 the frame PSNR moves from 35.55 to 35.35~dB and SSIM from 0.9194 to 0.9189; on Video~3 the changes are similarly marginal. The SQNR of 39.2~dB confirms that 16-bit rounding injects little noise relative to the signal.

Two effects deserve honest mention. First, on Video~3 the student's \emph{video} quality (28.86~dB) exceeds the teacher's (21.88~dB), but its \emph{audio} quality drops sharply (24.2~dB against the teacher's 62.5~dB). A plausible reading is that the capacity budget shifts toward video during distillation and the audio pays for it, though we have not isolated the cause. Second, audio PSNR on Video~1 \emph{improves} slightly after quantization (18.79 to 18.85~dB) while LSD worsens (4.59 to 4.85~dB); the rounding acts as weak regularization on amplitude error while blurring the spectrum. Neither effect changes the overall picture: the compression is nearly free at 16 bits.

\subsection{Quantization Bit-width Study}
\label{sec:quantstudy}

\begin{figure}[!t]
\centering
\begin{subfigure}[b]{0.49\columnwidth}
\includegraphics[width=\textwidth]{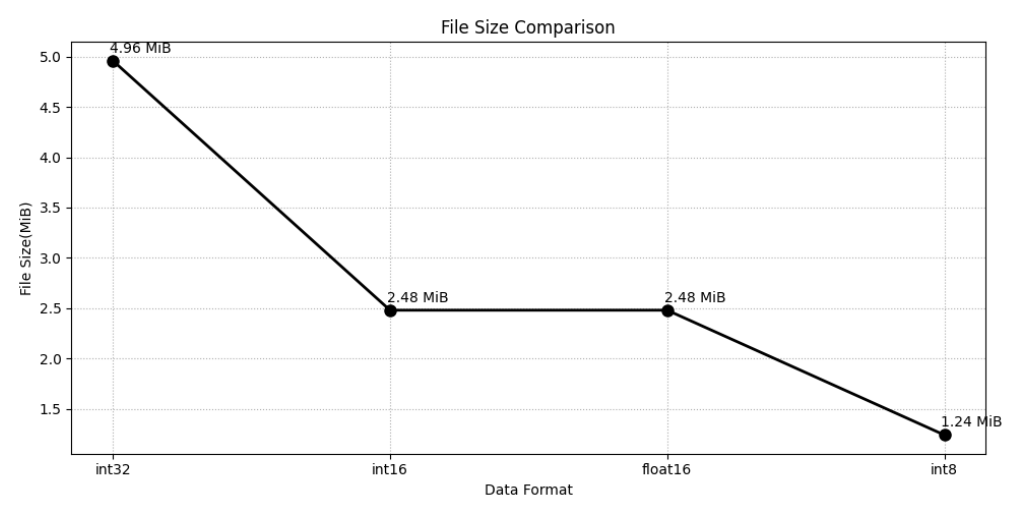}
\caption{File size.}
\end{subfigure}\hfill
\begin{subfigure}[b]{0.49\columnwidth}
\includegraphics[width=\textwidth]{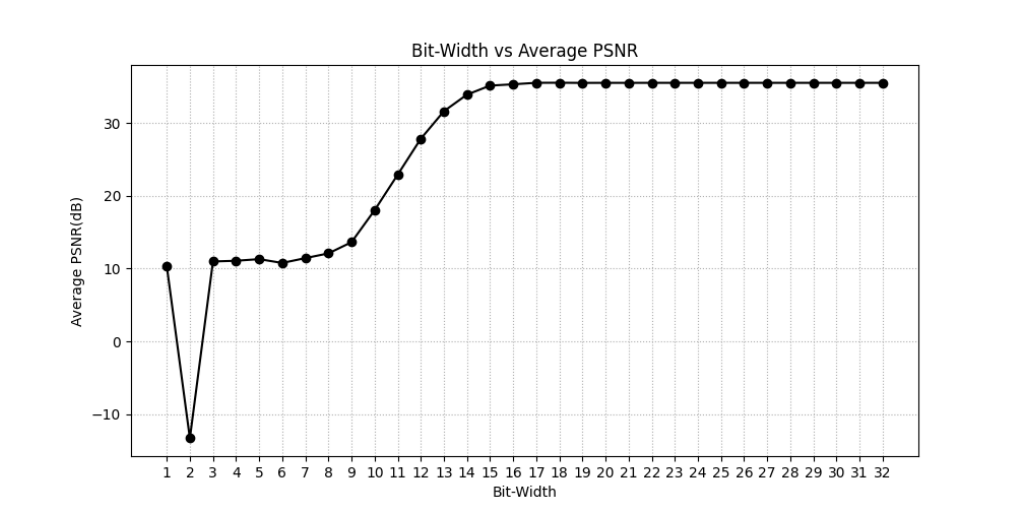}
\caption{Average PSNR.}
\end{subfigure}\\[2pt]
\begin{subfigure}[b]{0.49\columnwidth}
\includegraphics[width=\textwidth]{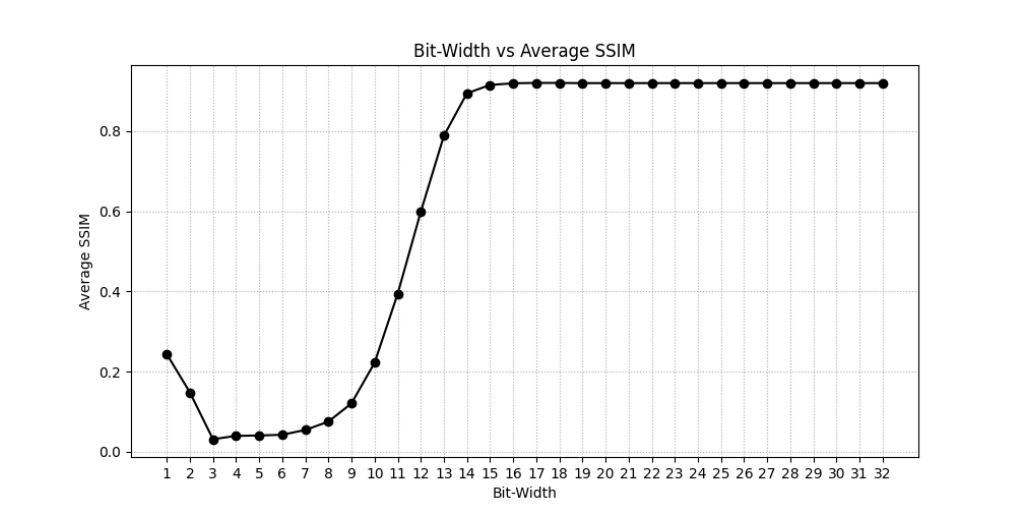}
\caption{Average SSIM.}
\end{subfigure}\hfill
\begin{subfigure}[b]{0.49\columnwidth}
\includegraphics[width=\textwidth]{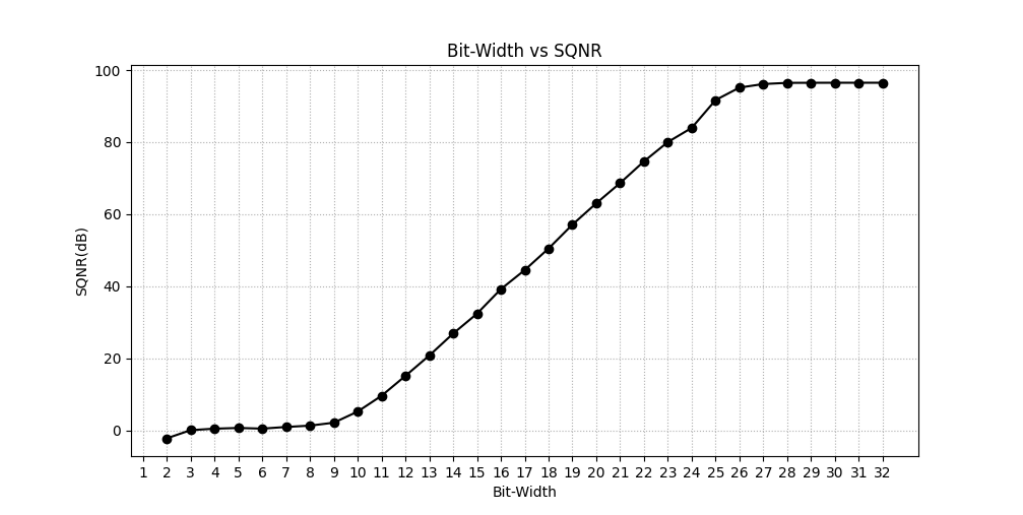}
\caption{SQNR.}
\end{subfigure}
\caption{Effect of quantization bit-width on the Video 1 student model. Quality metrics stabilize from 16 bits onward.}
\label{fig:quantsweep}
\end{figure}

\begin{figure}[!t]
\centering
\includegraphics[width=0.9\columnwidth]{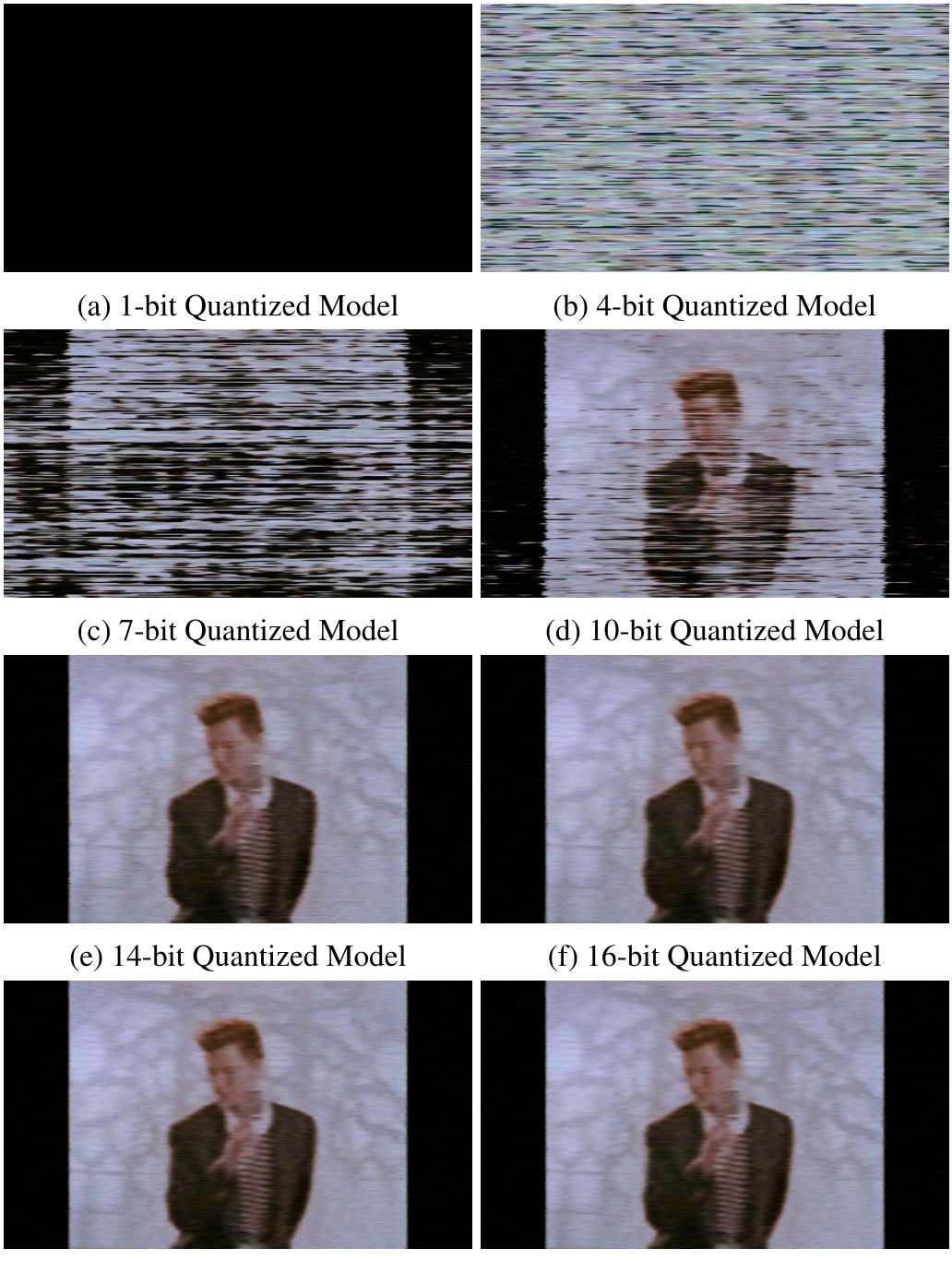}
\caption{Frame 10 of Video 1 reconstructed by the student model quantized to 1, 4, 7, 10, 14, 16, 18, and 32 bits.}
\label{fig:frames}
\end{figure}

To pick the operating bit-width we quantized the Video~1 student to every precision from int1 to int32 and measured average PSNR, SSIM, and SQNR over five reference frames (Fig.~\ref{fig:quantsweep}); file sizes could be exported for int8, int16, float16, and int32. Quality degrades severely at low bit-widths and recovers as precision grows, but from 16 bits onward the curves are flat --- more precision buys nothing visible. Figure~\ref{fig:frames} makes the same point with images: 1-bit and 4-bit outputs are unrecognizable, 7--14 bits progressively sharpen, and 16, 18, and 32 bits are indistinguishable to the eye. We therefore fix 16-bit quantization for the pipeline.

\begin{figure}[!t]
\centering
\includegraphics[width=0.92\columnwidth]{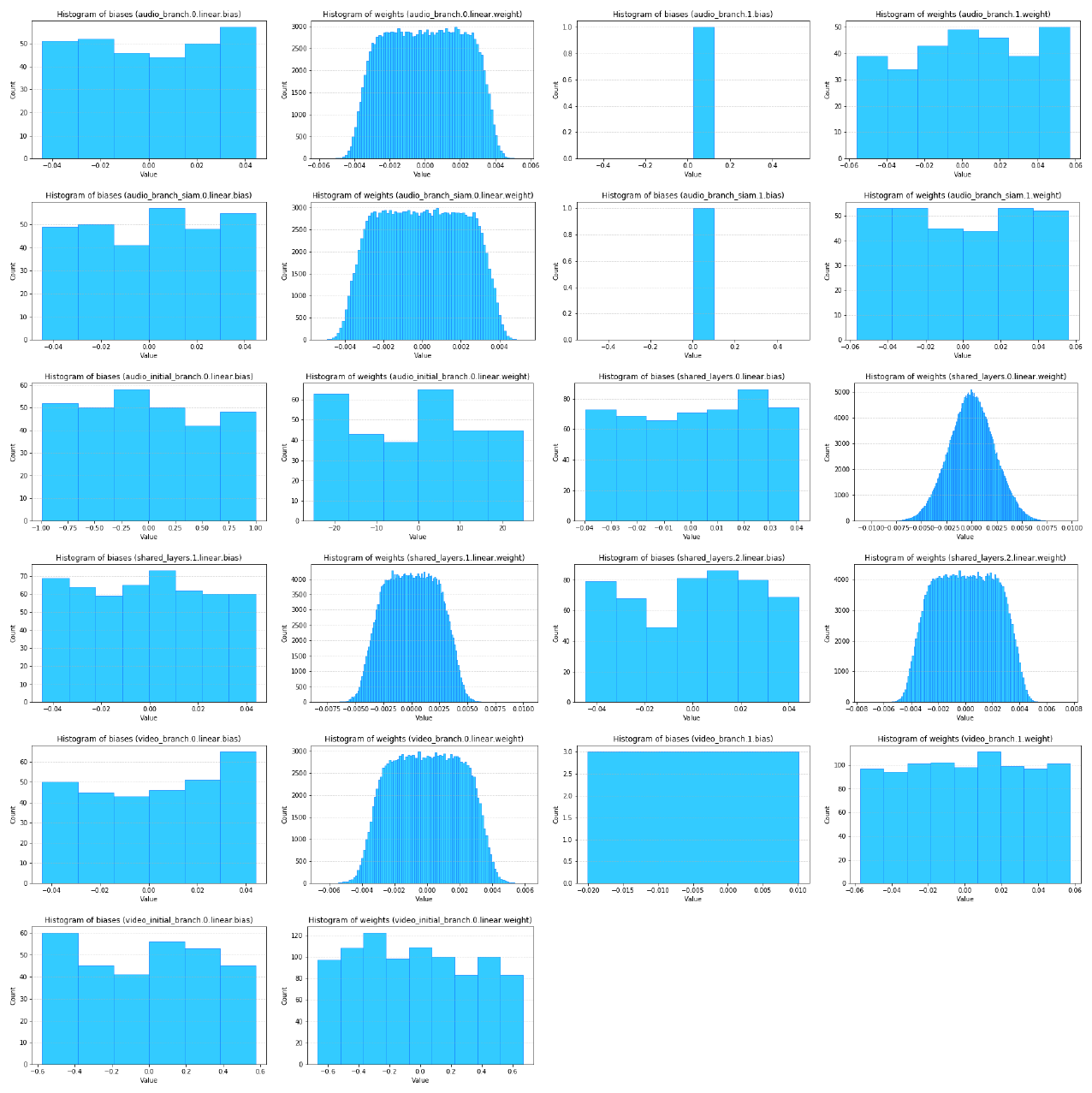}
\caption{Weight and bias histograms of the Video 1 student model, by layer group.}
\label{fig:hist}
\end{figure}

Figure~\ref{fig:hist} shows the student's weight and bias distributions by layer group. Weights are roughly Gaussian around zero --- friendly territory for symmetric quantization --- while biases are more uniform or lightly skewed.

\subsection{Lossless Encoding}

\begin{table}[!t]
\caption{Effect of LZMA2 (xz) Encoding on the Quantized Student}
\label{tab:encoding}
\centering
\footnotesize
\begin{tabular}{@{}lcccc@{}}
\toprule
& \multicolumn{2}{c}{Video 1} & \multicolumn{2}{c}{Video 3} \\
\cmidrule(lr){2-3}\cmidrule(l){4-5}
Metric & Before & After & Before & After \\
\midrule
File size (MiB)   & 2.48 & 2.43 & 2.48 & 2.33 \\
Compression ratio & 0.77 & 0.79 & 2.45 & 2.61 \\
\bottomrule
\end{tabular}
\end{table}

LZMA2 encoding is lossless, so quality metrics are untouched; Table~\ref{tab:encoding} shows the size effect. The gain is modest --- 2.48 to 2.43~MiB for Video~1 and 2.48 to 2.33~MiB for Video~3 --- because quantized network weights have little of the byte-level redundancy that dictionary coders exploit. Video~3 compresses slightly better, suggesting more redundancy in its weight stream. The final figure for Video~3 is an end-to-end compression ratio of 2.61 against the 6.08~MiB source.

\subsection{Comparison with Conventional Codecs}

\begin{table}[!t]
\caption{Video Metrics of Conventional Codecs, Video 1 (1{,}972 KiB)}
\label{tab:codecv1}
\centering
\footnotesize
\setlength{\tabcolsep}{3.5pt}
\begin{tabular}{@{}lccccccc@{}}
\toprule
Codec & CRF & \makecell{Bitrate\\(kbps)} & \makecell{PSNR\\(dB)} & SSIM & LPIPS & \makecell{Size\\(KiB)} & Ratio \\
\midrule
\multirow{3}{*}{H.264/MP3} & 1  & 320 & 61.60 & 0.99 & 0.004 & 429.64 & 4.59 \\
                           & 23 & 192 & 47.24 & 0.99 & 0.005 & 178.59 & 11.04 \\
                           & 51 & 64  & 32.00 & 0.83 & 0.23  & 44.28  & 44.52 \\
\midrule
\multirow{3}{*}{H.265/MP3} & 1  & 320 & 59.53 & 0.99 & 0.004 & 378.32 & 5.21 \\
                           & 28 & 192 & 45.32 & 0.98 & 0.02  & 154.51 & 12.76 \\
                           & 51 & 64  & 33.37 & 0.85 & 0.21  & 45.70  & 43.14 \\
\bottomrule
\end{tabular}
\end{table}

\begin{table}[!t]
\caption{Video Metrics of Conventional Codecs, Video 3 (6{,}235 KiB)}
\label{tab:codecv3}
\centering
\footnotesize
\setlength{\tabcolsep}{3.5pt}
\begin{tabular}{@{}lccccccc@{}}
\toprule
Codec & CRF & \makecell{Bitrate\\(kbps)} & \makecell{PSNR\\(dB)} & SSIM & LPIPS & \makecell{Size\\(KiB)} & Ratio \\
\midrule
\multirow{3}{*}{H.264/MP3} & 1  & 320 & 58.39 & 0.98 & 0.16 & 1{,}218.82 & 5.12 \\
                           & 23 & 192 & 42.22 & 0.96 & 0.16 & 161.85 & 38.52 \\
                           & 51 & 64  & 22.17 & 0.72 & 0.34 & 31.65  & 196.98 \\
\midrule
\multirow{3}{*}{H.265/MP3} & 1  & 320 & 56.07 & 0.99 & 0.16 & 861.80 & 7.23 \\
                           & 28 & 192 & 40.20 & 0.96 & 0.18 & 113.02 & 55.16 \\
                           & 51 & 64  & 24.16 & 0.73 & 0.35 & 31.42  & 198.39 \\
\bottomrule
\end{tabular}
\end{table}

\begin{table}[!t]
\caption{Audio Metrics of MP3, Videos 1 and 3}
\label{tab:mp3}
\centering
\footnotesize
\begin{tabular}{@{}ccccccc@{}}
\toprule
& \multicolumn{3}{c}{Video 1} & \multicolumn{3}{c}{Video 3} \\
\cmidrule(lr){2-4}\cmidrule(l){5-7}
\makecell{Bitrate\\(kbps)} & \makecell{PSNR\\(dB)} & LSD & ViSQOL & \makecell{PSNR\\(dB)} & LSD & ViSQOL \\
\midrule
64  & 35.34 & 1.97 & 4.66 & 27.99 & 4.21 & 4.64 \\
192 & 43.28 & 0.42 & 4.73 & 42.23 & 0.75 & 4.73 \\
320 & 64.78 & 0.10 & 4.73 & 66.45 & 0.18 & 4.73 \\
\bottomrule
\end{tabular}
\end{table}

We encoded all five videos with H.264 and H.265 (with MP3 audio) at three CRF settings each; Tables~\ref{tab:codecv1}--\ref{tab:mp3} show Videos~1 and~3, and the remaining videos follow the same pattern. The conventional codecs win on compression ratio by one to two orders of magnitude, which is the expected outcome against three decades of engineering.

The quality comparison is more nuanced. The teacher sits slightly below H.264/HEVC at CRF~23--28 on Videos~1, 2, and~5, and is comparable to their CRF~51 output on Videos~3 and~4. Its audio trails MP3 at 320~kbps on Videos~1--4 and matches MP3 at 192~kbps on Video~5. The quantized student matches the CRF~51 tier on Video~1 and slightly beats it on Video~3 (28.72~dB against 22.17--24.16~dB), while its audio lands near MP3 at 64~kbps. The INR codec is thus competitive only at the aggressive end of the conventional codecs' range --- but it reaches that point with a stored size that does not depend on video length or resolution, whereas conventional bitstreams scale linearly with content.

\subsection{Prototype Deployment}

The full pipeline runs behind a web interface. A FastAPI backend accepts an uploaded video, trains the INR in a subprocess, and returns the compressed model; browsers exchange models directly over WebRTC data channels, with WebSocket signaling and a public STUN server for NAT traversal, so transfers inside a network need no central storage. Received models are dequantized and inferenced server-side, and the reconstructed video plays back in the browser next to the original for visual comparison.

\section{Limitations and Future Work}

The measurements point at four concrete weaknesses. The fixed model-size floor (2.33--2.48~MiB after full compression) makes the codec useless for short clips and only mildly useful at the sizes we tested; the floor must drop, whether by smaller architectures, pruning, or entropy-constrained training of the kind proposed by Gomes et al.\ \cite{gomes2023entropy}. Dynamic content is the second weakness: reconstruction quality falls visibly as motion and frame rate increase, and representing high-fps video well remains open. Third, distillation currently trades audio quality for video quality; a loss schedule that protects the audio branch, or separate distillation temperatures per modality, is worth trying. Fourth, encoding means training a network per video for thousands of epochs, which rules out real-time use; meta-learned initializations or shared backbones across videos could cut this cost. On the quantization side, 16-bit symmetric rounding is simple but leaves room: quantization-aware training and mixed per-layer precision could push below 16 bits without the quality cliff we observed.

\section{Conclusion}

We built and measured a codec that stores video and audio jointly as the weights of one sine-activated MLP. Separate per-modality initialization layers turned out to be necessary --- audio wants first-layer weights in $(-25,25)$, video in $(-2/3,2/3)$ --- and a Siamese audio head provides a usable noise estimate for spectral denoising at decode time. Knowledge distillation, 16-bit symmetric quantization, and LZMA2 encoding together cut the stored representation from 9.05~MiB to 2.33~MiB while leaving reconstruction quality essentially unchanged, and a sweep over bit-widths shows 16 bits is the knee of the quality curve. Against H.264, HEVC, and MP3 the approach is competitive only at their lowest-quality settings, and its fixed size floor penalizes short inputs. What it offers in exchange is a single, differentiable, resolution-independent representation of both modalities whose stored size is constant in the content length --- a property no conventional codec has, and one that becomes more valuable as inputs grow. Lowering the size floor and handling motion are the clear next steps.

\balance


\begin{thebibliography}{15}
\footnotesize

\bibitem{kalva2006h264}
H.~Kalva, ``The H.264 video coding standard,'' \emph{IEEE MultiMedia}, vol.~13, no.~4, pp. 86--90, 2006.

\bibitem{pourazad2012hevc}
M.~T. Pourazad, C.~Doutre, M.~Azimi, and P.~Nasiopoulos, ``HEVC: The new gold standard for video compression: How does HEVC compare with H.264/AVC?'' \emph{IEEE Consumer Electronics Magazine}, vol.~1, no.~3, pp. 36--46, 2012.

\bibitem{sitzmann2020siren}
V.~Sitzmann, J.~N.~P. Martel, A.~W. Bergman, D.~B. Lindell, and G.~Wetzstein, ``Implicit neural representations with periodic activation functions,'' in \emph{Advances in Neural Information Processing Systems (NeurIPS)}, 2020.

\bibitem{chen2021nerv}
H.~Chen, B.~He, H.~Wang, Y.~Ren, S.-N. Lim, and A.~Shrivastava, ``NeRV: Neural representations for videos,'' in \emph{Advances in Neural Information Processing Systems (NeurIPS)}, 2021.

\bibitem{gomes2023entropy}
C.~Gomes, R.~Azevedo, and C.~Schroers, ``Video compression with entropy-constrained neural representations,'' in \emph{Proceedings of the IEEE/CVF Conference on Computer Vision and Pattern Recognition (CVPR)}, 2023, pp. 18\,497--18\,506.

\bibitem{zhang2024boosting}
X.~Zhang, R.~Yang, D.~He, X.~Ge, T.~Xu, Y.~Wang, H.~Qin, and J.~Zhang, ``Boosting neural representations for videos with a conditional decoder,'' 2024, arXiv:2402.18152.

\bibitem{lanzendorfer2023siamese}
L.~A. Lanzend\"orfer and R.~Wattenhofer, ``Siamese SIREN: Audio compression with implicit neural representations,'' 2023, arXiv:2306.12957.

\bibitem{choudhury2024nerva}
A.~Choudhury, P.~Singh, and G.-M. Su, ``NeRVA: Joint implicit neural representations for videos and audios,'' in \emph{2024 IEEE International Conference on Multimedia and Expo (ICME)}, 2024, pp. 1--6.

\bibitem{shlien1994mpeg}
S.~Shlien, ``Guide to MPEG-1 audio standard,'' \emph{IEEE Transactions on Broadcasting}, vol.~40, no.~4, pp. 206--218, 1994.

\bibitem{li2021dcvc}
J.~Li, B.~Li, and Y.~Lu, ``Deep contextual video compression,'' in \emph{Advances in Neural Information Processing Systems (NeurIPS)}, 2021.

\bibitem{li2023diverse}
J.~Li, B.~Li, and Y.~Lu, ``Neural video compression with diverse contexts,'' in \emph{Proceedings of the IEEE/CVF Conference on Computer Vision and Pattern Recognition (CVPR)}, 2023.

\bibitem{liu2021latent}
B.~Liu, Y.~Chen, S.~Liu, and H.-S. Kim, ``Deep learning in latent space for video prediction and compression,'' in \emph{Proceedings of the IEEE/CVF Conference on Computer Vision and Pattern Recognition (CVPR)}, 2021, pp. 701--710.

\bibitem{panneerselvam2022effective}
K.~Panneerselvam, K.~Mahesh, V.~Josephine, and R.~Anandan, ``Effective and efficient video compression by the deep learning techniques,'' \emph{Computer Systems Science and Engineering}, vol.~45, pp. 1047--1061, 2022.

\bibitem{han2024kdinr}
J.~Han, H.~Zheng, and C.~Bi, ``KD-INR: Time-varying volumetric data compression via knowledge distillation-based implicit neural representation,'' \emph{IEEE Transactions on Visualization and Computer Graphics}, vol.~30, no.~10, pp. 6826--6838, 2024.

\bibitem{pytorch_quant}
PyTorch, ``Quantization --- PyTorch documentation,'' \url{https://pytorch.org/docs/stable/quantization.html}, accessed 2024-10-15.

\end{thebibliography}
\end{document}